\documentclass[british,preprint,prb,aps]{revtex4-1}
\usepackage[T1]{fontenc}
\usepackage[latin9]{inputenc}
\usepackage{amsmath}
\usepackage{amssymb}
\usepackage{graphicx}

\usepackage{color}
\usepackage[normalem]{ulem}

\newcommand{\red}[1]{#1}
\newcommand{\blue}[1]{}



\usepackage{babel}
\begin{document}
\global\long\def\non{}
\newcommand{\difp}[2]{\frac{\partial #1}{\partial #2}}
\newcommand{\be}[1]{\begin{equation} \label{#1}}
\newcommand{\ee}{\end{equation}}
\newcommand{\bea}[1]{\begin{eqnarray} \label{#1}}
\newcommand{\eea}{\end{eqnarray}}
\newcommand{\bean}{\begin{eqnarray*}}
\newcommand{\eean}{\end{eqnarray*}}
\global\long\def\br{{\bf r}}
\global\long\def\bR{{\bf R}}
\global\long\def\bA{{\bf A}}
\global\long\def\bB{{\bf B}}
\global\long\def\bE{{\bf E}}
\global\long\def\bm{{\bf m}}
\global\long\def\bn{{\bf n}}
\global\long\def\bN{{\bf N}}
\global\long\def\bp{{\bf p}}
\global\long\def\bF{{\bf F}}
\global\long\def\bz{{\bf z}}
\global\long\def\bZ{{\bf Z}}
\global\long\def\bV{{\bf V}}
\global\long\def\bv{{\bf v}}
\global\long\def\bu{{\bf u}}
\global\long\def\bx{{\bf x}}
\global\long\def\bX{{\bf X}}
\global\long\def\bJ{{\bf J}}
\global\long\def\bj{{\bf j}}
\global\long\def\bk{{\bf k}}
\global\long\def\bTheta{{\bf \Theta}}
\global\long\def\btheta{\boldsymbol{\theta}}
\global\long\def\bOmega{\boldsymbol{\Omega}}
\global\long\def\bomega{\boldsymbol{\omega}}
\global\long\def\brho{\boldsymbol{\rho}}
\global\long\def\rd{{\rm d}}
\global\long\def\rJ{{\rm J}}
\global\long\def\ph{\varphi}
\global\long\def\te{\theta}
\global\long\def\tht{\vartheta}
\global\long\def\vparkb{v_{\parallel k b}}
\global\long\def\vparkm{v_{\parallel k m}}
\global\long\def\Jpar{J_{\parallel}}
\global\long\def\ppar{p_{\parallel}}
\global\long\def\Bpstar{B_{\parallel}^{*}}
\global\long\def\intpi{\int\limits _{0}^{2\pi}}
\global\long\def\summ{\sum\limits _{m=-\infty}^{\infty}}
\global\long\def\tb{\tau_{b}(\uv)}
\global\long\def\bh{{\bf h}}
\global\long\def\cE{{\cal E}}
\global\long\def\odtwo#1#2{\frac{\rd#1}{\rd#2}}
\global\long\def\pdone#1{\frac{\partial}{\partial#1}}
\global\long\def\pdtwo#1#2{\frac{\partial#1}{\partial#2}}
\global\long\def\ds{{\displaystyle }}
\global\long\def\iotabar{\mbox{\ensuremath{\iota\!\!}-}}
\global\long\def\vpar{v_{\parallel}}

\renewcommand{\be}[1]{\begin{equation} \label{#1}}
\renewcommand{\ee}{\end{equation}}
\renewcommand{\bea}[1]{\begin{eqnarray} \label{#1}}
\renewcommand{\eea}{\end{eqnarray}}
\newcommand{\eq}[1]{(\ref{#1})}

\author{Christopher~G.~Albert$^1$, Martin~F.~Heyn$^1$, Gernot~Kapper$^1$, Sergei~V.~Kasilov$^{1,2}$,  Winfried~Kernbichler$^1$, Andreas~F.~Martitsch$^1$}
\affiliation{$^1$Fusion@\"OAW, Institut f\"ur Theoretische Physik - Computational Physics,\\  Technische Universit\"at Graz, Petersgasse~16, 8010 Graz, Austria \\ $^2$Institute of Plasma Physics, National Science Center ``Kharkov Institute of Physics and\\ Technology'',  ul.~Akademicheskaya 1, 61108 Kharkov, Ukraine}

\title{Evaluation of toroidal torque by non-resonant magnetic perturbations
in tokamaks for resonant transport regimes using a Hamiltonian approach}
\begin{abstract}
Toroidal torque generated by neoclassical viscosity caused by external
non-resonant, non-axisymmetric perturbations has a significant influence
on toroidal plasma rotation in tokamaks. In this article, a derivation
for the expressions of toroidal torque and radial transport in resonant
regimes is provided within quasilinear theory in canonical action-angle
variables. The proposed approach treats all low-collisional 
quasilinear resonant NTV regimes including superbanana plateau
and drift-orbit resonances in a unified way and allows for magnetic drift 
in all regimes.
It is valid for perturbations on toroidally symmetric flux surfaces of the 
unperturbed equilibrium without specific assumptions on geometry or aspect ratio. 
The resulting expressions are shown to match existing analytical results in the
large aspect ratio limit. Numerical results from the newly developed code NEO-RT are 
compared to calculations by the quasilinear version of the code NEO-2 at 
low collisionalities.
The importance of the magnetic shear term in the magnetic drift frequency 
and a significant effect of the magnetic drift on drift-orbit resonances 
are demonstrated.
\end{abstract}
\maketitle

\section{Introduction}

In tokamaks, non-axisymmetric magnetic field perturbations such as toroidal field ripple, 
error fields and perturbation fields from Edge Localized Mode (ELM) mitigation coils
produce non-ambipolar radial transport at non-resonant flux surfaces occupying most of the plasma volume. 
The toroidal torque associated with this transport significantly changes the toroidal plasma rotation -- 
an effect known as neoclassical toroidal 
viscosity~\cite{zhu06-225002,shaing09-075015,Park2009-65002,shaing10-025022,kasilov14-092506,Shaing2015-125001} 
(NTV). 
At low collisionalities,
resonant transport regimes~\cite{yushmanov82-1123,yushmanov90-117}, 
namely superbanana plateau~\cite{shaing09-035009,shaing15-905810203},
bounce and bounce-transit (drift-orbit) resonance regimes~\cite{shaing09-075015},
have been found to play an important role in modern tokamaks, in particular 
in ASDEX Upgrade\red{~\cite{martitsch16-074007}}.
In these regimes, which emerge if perturbation field amplitudes are small enough,
transport coefficients become independent of the collision frequency (form a plateau).
The interaction of particles with the (quasi-static) electromagnetic field in these plateau-like regimes 
is a particular case of collisionless wave-particle interaction with time dependent fields
and can be described within quasilinear theory. The most compact form of this theory in application to a
tokamak geometry is obtained in canonical action-angle 
variables~\cite{Kaufman1972-1063,Hazeltine1981-1164,Mahajan1983-700,becoulet91-137,timofeev94-336,kominis08}.
Here, this formalism is applied to ideal quasi-static electromagnetic perturbations, 
which can be described in terms of flux coordinates. As a starting point, the Hamiltonian description
of the guiding center motion in those coordinates in general 3D magnetic 
configurations (see, e.g., Refs.~\onlinecite{White1982-575,White1984-2455,White1990-845}) is used.
For the particular case of Boozer coordinates the perturbation theory is constructed with 
respect to non-axisymmetric perturbations of the magnetic field module, which is the only function of 
angles relevant for neoclassical transport.

\noindent
The purpose of this paper is twofold: The first aim is to describe the NTV in all quasilinear 
resonant regimes in a unified form using the standard Hamiltonian formalism and to develop
a respective numerical code allowing for fast NTV evaluation in these regimes without any
simplifications to the magnetic field geometry. The second aim is to benchmark this approach
with the quasilinear version of the NEO-2 
code\red{~\cite{kasilov14-092506,martitsch16-074007}}
which treats the general case of plasma collisionality. 
Since particular resonant regimes described in
literature basically agree with the Hamiltonian approach within their applicability domains,
such a benchmarking means also the benchmarking of NEO-2 against those results.
The structure of the paper is as follows. In section~\ref{sec:transeqs}, basic definitions are given
and two different quasilinear expressions for the toroidal torque density are derived for the general 
case of small amplitude quasi-static electromagnetic perturbations. 
In section~\ref{sec:tokamak} the perturbation theory for ideal perturbations described by small corrugation
of magnetic surfaces in flux coordinates is outlined, and expressions for the canonical action-angle
variables are given. In section~\ref{sec:neotorque} expressions for non-axisymmetric transport coefficients
are derived, and in section~\ref{sec:numerics} the numerical implementation of the Hamiltonian formalism in
the code NEO-RT is presented and its results compared with the results of NEO-2 code for typical resonant transport regimes.
The results are summarized in section~\ref{sec:conclosion}.

\section{Transport equations and toroidal torque in Hamiltonian variables}
\label{sec:transeqs}

\noindent
In Hamiltonian variables the kinetic equation can be compactly written in the form
\be{kineqnonlin}
\difp{f}{t}+\left\{f,H\right\}=\hat L_c f,
\ee
where $\hat L_c$ is the collision operator and
\be{pois}
\left\{f,g\right\} \equiv 
\difp{f}{\br}\cdot\difp{g}{\bp}-\difp{f}{\bp}\cdot\difp{g}{\br}
=
\difp{f}{\theta^i}\difp{g}{J_i}-\difp{f}{J_i}\difp{g}{\theta^i}
=
\difp{}{\theta^i}\left(f\difp{g}{J_i}\right)
-
\difp{}{J_i}\left(f\difp{g}{\theta^i}\right)
\ee
is the Poisson bracket which is invariant with respect to the canonical variable choice. Here,
$(\br,\bp)$ are Cartesian coordinates and canonical momentum components, $(\btheta,\bJ)$ are canonical angles and actions specified later, summation over repeated indices is assumed,
and bold face describes a whole set of three variables (e.g. $\btheta=(\theta^1,\theta^2,\theta^3)$).
In the following derivations, straight field line flux coordinates $\bx=(r,\vartheta,\varphi)$ are used with a specific definition of
the flux surface label (effective radius) such that $\langle|\nabla r|\rangle=1$,
where the neoclassical magnetic flux surface average is given by
\be{fluxav}
\langle a \rangle=\frac{1}{S}\int\limits_{-\pi}^\pi \rd \vartheta\int\limits_{-\pi}^\pi \rd \varphi\;\sqrt{g}\; a,
\qquad
S=\int\limits_{-\pi}^\pi \rd \vartheta\int\limits_{-\pi}^\pi \rd \varphi\;\sqrt{g},
\ee
and $\sqrt{g}$ is the metric determinant. 
Due to the above definition of $r$, quantity $S$ has the meaning of the flux surface area.

\noindent
Multiplying~\eq{kineqnonlin} by a factor $a\,\delta\left(r-r_c\right)$ where 
$a=a(\btheta,\bJ)=a(\br,\bp)$ is some function of particle position in the phase space
and $r_c=r_c(\btheta,\bJ)=r(\br_c(\btheta,\bJ))$ is the particle effective radius 
expressed via phase space variables, integrating over the phase space and 
dividing the result by the flux surface area $S$ leads to a generalized conservation law 
\be{conslaw}
\difp{A}{t}+\frac{1}{S}\difp{}{r}S\Gamma_A=s_A+s_{A}^{(c)},
\ee
where
\bea{pphidens}
A=A(t,r)
&\equiv&
\frac{1}{S} \int\rd^3 \theta \int\rd^3 J\; \delta\left(r-r_c\right)  a f
=
\frac{1}{S}\int\rd^3 r \delta\left(r-r_c\right) \int\rd^3 p\;  a f
\nonumber \\
&=&\frac{1}{S}\int\limits_{-\pi}^\pi \rd \vartheta \int\limits_{-\pi}^\pi \rd \varphi \;
\sqrt{g}\int\rd^3 p\;  a f
=\left\langle \int\rd^3 p\; a f \right\rangle,
\eea
where $\delta(\dots)$ is the Dirac delta function. Generalized magnetic surface averaged flux and source densities in~\eq{conslaw} are given, respectively, by
\bea{fluxsource}
\Gamma_A
&\equiv&
\frac{1}{S}\int\rd^3 \theta \int\rd^3 J\; \delta\left(r-r_c\right)  \left\{r_c,H\right\} a f,
\\
s_A
&\equiv&
\frac{1}{S}\int\rd^3 \theta \int\rd^3 J\; \delta\left(r-r_c\right)  \left\{a,H\right\} f,
\eea
where the second representation of the Poisson bracket~\eq{pois} has been used for these expressions,
and the collisional source density is
\be{nocolint}
s_{A}^{(c)}=\left\langle\int\rd^3 p\; a \hat L_c f\right\rangle.
\ee
For $a=1$ the continuity equation is obtained with no sources, $s_n=s_n^{(c)}=0$
and surface averaged particle flux density $\Gamma_A=\Gamma$ given by
\be{partflux}
\Gamma = \frac{1}{S} \int\rd^3 \theta \int\rd^3 J\; \delta\left(r-r_c\right)  \left\{r_c,H\right\} f.
\ee
For $a=p_\varphi$ with 
\be{pphi}
p_\varphi=\bp\cdot\difp{\br}{\varphi}=m_\alpha v_\varphi + \frac{e_\alpha}{c}A_\varphi(r)
\ee
being the canonical angular momentum, the
equation for the canonical angular momentum density is obtained
with the source term $s_a=s_{p_\varphi}=T_\varphi^{\rm NA}$ being the toroidal torque density acting on the
given species from the electromagnetic field,
\be{torque}
T_\varphi^{\rm NA} = -\frac{1}{S}\int\rd^3 \theta \int\rd^3 J\; \delta\left(r-r_c\right)  \difp{H}{\varphi} f.
\ee
In Eq.~\eq{pphi}, $v_\varphi=\bv\cdot \partial\br/\partial\varphi$ and $A_\varphi=-\psi_{\rm pol}$ 
are covariant toroidal velocity and vector potential components, respectively and 
$\psi_{\rm pol}$ is the normalized poloidal flux. In addition, speed of light $c$, 
and charge $e_\alpha$ and mass $m_\alpha$ of species $\alpha$ appear in the expression. 

\noindent
One can see that torque density is determined only by the non-axisymmetric part of the distribution function
while the particle flux density contains also the axisymmetric contribution. This property of the torque
is rather helpful in the nonlinear transport theory which, however, is not the topic of the present paper.
A conservation law of the kinematic toroidal momentum, $a=m_\alpha v_\varphi$,
is obtained by subtraction of the continuity equation multiplied by $e_\alpha A_\varphi /c$. 
The source term in this equation is
\be{kinmomsource}
s_{m_\alpha v_\varphi}=T_\varphi^{\rm NA}+s_{p_\varphi}^{(c)}+\frac{e_\alpha}{c}\sqrt{g}B^\vartheta \Gamma,
\ee
where $B^\vartheta$ is the poloidal contravariant magnetic field component.
Assuming a static momentum balance and estimating $\Gamma_{m_\alpha v_\varphi} \sim m_\alpha v_\varphi \Gamma$,
which means that contribution of the radial momentum transport term to this balance is negligible 
because it scales to the last term in~\eq{kinmomsource} as $q\rho_L R/r^2 \ll 1$,
this balance is reduced to $s_{m_\alpha v_\varphi}=0$. Here $q$, $\rho_L$, $R$ and $r$ 
are safety factor, Larmor radius, major and minor radius, respectively.
The result is a flux-force relation~\cite{hirshman78-917,shaing83-3315}, 
which links particle flux to the torques (a static density equilibrium without particle 
sources where $\Gamma=0$ demonstrates the fact that $T_\varphi^{\rm NA}$ is indeed a torque density 
because it balances collisional momentum source density $s_{p_\varphi}^{(c)}$ alone). 
The presence of the collisional force moment $s_{p_\varphi}^{(c)}$ in the flux force relation
indicates that the calculation of torque and radial flux needs a certain caution when using 
a Krook collision model, which is usually the case in quasilinear ``collisionless'' plateau 
transport regimes described here.
Due to momentum conservation by collisions, collisional torque $s_{p_\varphi}^{(c)}$ provides no 
contribution to the total torque that is of main interest here, which is not ensured by the simple 
Krook model.
This is the case, in particular, for the ion component in the simple plasma where momentum is largely
conserved within this component. Thus, when computing particle flux density in this case, one should keep
in mind that direct computation of $\Gamma$ from the quasilinear equation provides a different result
as compared to such computation through $T_\varphi^{\rm NA}$ via the flux force relation with no collisional 
torque $s_{p_\varphi}^{(c)}$,
\be{eq:FluxForce}
T_{\ph}^{\rm NA}=-\frac{e_\alpha}{c}\sqrt{g}B^\vartheta \Gamma=
-\frac{e_{\alpha}}{c}\frac{\rd \psi_{{\rm pol}}}{\rd r} \Gamma.
\ee
Since $T_\varphi^{\rm NA}$ is not affected by details of the collision model,
this more appropriate definition of $\Gamma$ is assumed below unless otherwise mentioned.
\red{
It should be noted that in the standard neoclassical theory~\cite{hinton76-239}
momentum conservation terms
are usually treated first before any approximations on the collision operator
are made thus avoiding the errors of the kind discussed above.
}

\noindent
Further steps are standard for quasilinear theory in action-angle variables~\cite{Kaufman1972-1063}. One presents the Hamiltonian and the distribution 
function as a sum of the unperturbed part depending on actions only and a perturbation with zero average over canonical angles,
$H(\btheta,\bJ)=H_0(\bJ)+\delta H(\btheta,\bJ)$ and $f(\btheta,\bJ)=f_0(\bJ)+\delta f(\btheta,\bJ)$, respectively and expands
the perturbations into a Fourier series over canonical angles,
\be{expser}
\delta H(\btheta,\bJ)=\sum_\bm H_\bm(\bJ){\rm e}^{im_k\theta^k},
\qquad
\delta f(\btheta,\bJ)=\sum_\bm f_\bm(\bJ){\rm e}^{im_k\theta^k},
\ee
where sums exclude $\bm=(0,0,0)$ term. By using a Krook collision term with infinitesimal collisionality,
$\hat L_c f = -\nu \delta f\rightarrow 0$,
the amplitudes of the perturbed distribution function from the linear order equation follow as
\be{linorder}
\left\{\delta f,H_0\right\}+\left\{f_0,\delta H\right\}+\nu \delta f = 
\sum_\bm\left(
\left(im_k\Omega^k+\nu\right)f_\bm-iH_\bm m_k\difp{f_0}{J_k}
\right) {\rm e}^{im_k\theta^k}=0.
\ee
Here, $\Omega^{k}=\partial H/\partial J_{k}$ are canonical frequencies, and the
time derivative has been omitted as small compared to all canonical frequencies in case of quasi-static
perturbations of interest here. A quasilinear equation is obtained by retaining only secular,
angle-independent terms in the second order equation,
\be{eq:quasilinear}
\difp{f_0}{t}+\overline{\left\{\delta f,\delta H\right\}}
 =
\difp{f_0}{t}-\sum_\bm m_k\difp{Q_\bm}{J_k}=0,
\ee
where the over-line stands for the average over the angles, and
\be{eq:Qm}
Q_{\bm}=Q_{\bm}(\bJ)=\frac{\pi}{2}|H_{\bm}|^{2}\delta(m_{j}\Omega^{j})m_{k}\difp{f_{0}}{J_{k}}
\ee
contains a resonance condition in the argument of a delta function that follows from the limit $\nu\rightarrow 0$.
The knowledge of $f_\bm$ is already sufficient for the evaluation of torque densities from Eq.~\eq{torque} where
the derivative over $\varphi$ is equivalent to a derivative over the canonical toroidal phase $\theta^3$,
\be{torque_ql}
T^{\rm NA}_{\ph}=-\frac{m_3}{S}\int d^{3}\theta\int d^{3}J\delta(r-r_c)\sum_\bm Q_{\mathbf{m}},
\ee
and of the particle flux from~\eq{partflux} 
\be{partlux_ql}
\Gamma_{\rm F}=-\frac{1}{S}\int d^{3}\theta\int d^{3}J\delta(r-r_c)\sum_{\mathbf{m}}m_{k}
\difp{r_{c}}{J_k}Q_{\mathbf{m}},
\ee
which is distinguished here from~\eq{eq:FluxForce} by subscript $F$. 
Alternatively the same expressions are obtained computing the conservation laws using
the quasilinear equation~\eq{eq:quasilinear} as a starting point~\cite{Hazeltine1981-1164}.
If the collision model does not conserve the parallel momentum such as e.g. the Krook model,
direct calculations of the torque in terms of viscosity~\cite{shaing09-075015}
and calculation of the torque through particle flux~\cite{Park2009-65002} using the force-flux 
relation~\eq{eq:FluxForce} may lead to different results. This difference, however, is
negligible in resonant transport regimes where details of the collision model are not important,
and collisionality can be treated as infinitesimal.

\section{Tokamak with ideal non-axisymmetric quasi-static perturbations}
\label{sec:tokamak}

\subsection{Canonical Hamiltonian variables for perturbed equilibria}

\noindent
Often in quasilinear theory in action-angle variables, both, the unperturbed and perturbed
Hamiltonian correspond to physically possible motion with separation of the unperturbed 
electromagnetic field and its perturbation in real space. However, there is no mathematical need
to do so. In particular, if the perturbed equilibrium is ideal such that it can be described
in flux coordinates, it is more convenient to restrict the perturbations only to those
quantities in the Hamiltonian which violate the axial symmetry. In case of Boozer coordinates
and also in many cases described in Hamada coordinates
the only important quantity is the magnetic field module which is generally
adopted for the construction of perturbation theory for NTV 
models~\cite{shaing09-075015,Park2009-65002,shaing10-025022,kasilov14-092506,shaing15-905810203}.
Thus the guiding center Lagrangian~\cite{Littlejohn1983-111} is transformed here to flux coordinates 
$\bx=(r,\vartheta,\varphi)$ as a starting point,
\be{gclagr}
L=m_{\alpha}v_{\parallel}h_{r}\dot{r}
+(m_{\alpha}v_{\parallel}h_{\tht}+\frac{e_\alpha}{c}A_{\tht})\dot{\tht}
+(m_{\alpha}v_{\parallel}h_{\ph}+\frac{e_{\alpha}}{c}A_{\ph})\dot{\ph}
+J_{\perp}\dot{\phi}-H,
\ee
where lower subscripts denote covariant components (in particular, 
$A_\vartheta=A_\vartheta(r)=\psi_{\rm tor}$ is the covariant poloidal component of the vector potential,
which is equal to the normalized toroidal flux and $A_r=0$), 
$\mathbf{h}=\mathbf{B}/B$ is the unit vector along the magnetic field,
$v_{\parallel}$ is the parallel velocity, $J_{\perp}=m_{\alpha}v_{\perp}^{2}/(2\omega_{c})$
is the perpendicular adiabatic invariant with $v_\perp$ and $\omega_{c}$ being the
perpendicular velocity and cyclotron frequency, respectively, $\phi$ is the gyrophase and
the Hamiltonian is given explicitly below in Eq.~\eq{expandham}.
The canonical form of the Lagrangian is obtained by transforming the toroidal angle 
$\ph$ to
\be{modvarphi}
\ph_{H} =\ph-\frac{c m_{\alpha}v_{\parallel}h_{r}}{e_\alpha A_{\ph}^{\prime}},
\ee
where the prime stands for a radial derivative. 
Omitting a total time derivative,
the Lagrangian transforms to 
\be{}
L=p_\vartheta \dot{\tht}
+p_\varphi\dot{\ph}_{H}
+J_{\perp}\dot{\phi}-H+
\frac{c m_{\alpha}^2 v_{\parallel}h_{\ph}}{e_\alpha}
\frac{d}{dt}\left(\frac{v_{\parallel}h_{r}}{A_{\ph}^{\prime}}\right),
\ee
where 
\be{momenta}
p_\vartheta=m_{\alpha}v_{\parallel}h_{\tht}(\bx)+\frac{e_{\alpha}}{c}A_{\tht}(r),
\qquad
p_\varphi=m_{\alpha}v_{\parallel}h_{\ph}(\bx)+\frac{e_{\alpha}}{c}A_{\ph}(r)
\ee
are canonical momenta in guiding center approximation, and
the last term is of the next order in $\rho_{\parallel}=v_{\parallel}/\omega_{c}$
and should therefore be neglected. 
Transformation~\eq{modvarphi} affects only
a small non-axisymmetric part of the field and
is different from the one of Refs.~\cite{White1984-2455,White1990-845} where the          
poloidal angle $\tht$ is modified instead. 
Alternatively, for collisionless transport regimes of interest here, 
one can simply ignore the covariant magnetic field component $B_r$ because it
does not contribute to the radial guiding center velocity, and its contribution
to the rotation velocity vanishes on a time scale larger than bounce time.

\noindent
Since the momenta are the independent variables, Eq.~\eq{momenta} should be regarded as
a definition of $r$ and $v_\parallel$. For the construction of perturbation theory
in Boozer coordinates being the main choice here,
the last quantity is redefined via the unperturbed parallel velocity $v_{0\parallel}$ as follows,
\be{vparunpert}
v_\parallel = v_{0\parallel}\frac{B(\bx)}{B_0(r,\vartheta)}
\ee
where subscript 0 corresponds to the axisymmetric part of the respective quantity.
Due to such redefinition, $r$ and $v_{0\parallel}$ do not depend on the toroidal angle $\varphi$
because in Boozer coordinates this dependence vanishes in both expressions in~\eq{momenta} 
due to $h_{\vartheta,\varphi}=B_{\vartheta,\varphi}(r)/B(\bx)$.
For the comparison with the results obtained in Hamada coordinates for the superbanana-plateau
regime, which is a resonant regime described by the bounce-averaged equation,
the definition of the unperturbed parallel velocity is opposite to~\eq{vparunpert},
$v_\parallel = v_{0\parallel} B_0/B$. With this redefinition, angular covariant components of $\bh$
in~\eq{vparunpert} are transformed within linear order in the perturbation field as follows,
\be{transcovham}
B h_k = B_0 h_{0k}
+\difp{\delta \chi}{x^k}-h_{0k}h_0^j \difp{\delta \chi}{x^j}, \qquad k=2,3,
\ee
where $\delta \chi$ is the non-axisymmetric perturbation of a function $\chi$ which enters the
definition of co-variant magnetic field components in Hamada coordinates $B_k$ via their flux surface 
averages $\bar B_k=\bar B_k(r)$, with $B_k=\bar B_k+\partial \chi/\partial x^k$.
Terms with $\delta \chi$, whose contribution in~\eq{transcovham} is orthogonal to the unperturbed
magnetic field, can be simply ignored in bounce-averaged regimes because they do not contribute to
bounce averaged velocity components.

\noindent
Thus, the Hamiltonian is expanded in Boozer coordinates up to a linear 
order in the perturbation field amplitude as follows,
\be{expandham}
H=\omega_c J_\perp+\frac{m_\alpha v_\parallel^2}{2}+e_\alpha\Phi=
\frac{B}{B_0}\omega_{c0} J_\perp+\frac{B^2}{B_0^2}\frac{m_\alpha v_{0\parallel}^2}{2}+e_\alpha\Phi
\approx H_0+\delta H,
\ee
where $\Phi=\Phi(r)$ is the electrostatic potential, 
\be{H0deltaH}
H_0=\omega_{c0} J_\perp+\frac{m_\alpha v_{0\parallel}^2}{2}+e_\alpha\Phi,
\qquad
\delta H=\left(\omega_{c0} J_\perp+m_\alpha v_{0\parallel}^2\right)\frac{\delta B}{B_0}.
\ee
The Hamiltonian perturbation $\delta H$ in Hamada coordinates differs from~\eq{H0deltaH}
by the opposite sign of the second term in the parentheses, $m_\alpha v_{0\parallel}^2$.
This term is usually ignored in tokamaks with large aspect ratio $A$ because for trapped
and barely trapped particles which are mainly contributing to NTV at small Mach numbers
(at sub-sonic toroidal rotation velocities) it scales to the first term as $1/A$.

\subsection{Action-angle variables in the axisymmetric tokamak}
\label{ssec:canvar}

\noindent
Since this subsection deals only with unperturbed motion corresponding to $H=H_0$,
the subscript 0 is dropped on all quantities here which are strictly axisymmetric.
Here it is convenient to replace the toroidal momentum $p_{\ph}$, which is
now a conserved quantity, by another invariant of motion $r_\varphi$ which describes
the banana tip radius for trapped particles~\cite{Hazeltine1981-1164} and is implicitly 
defined via
\be{eq:pphi}
\frac{e_{\alpha}}{c}A_{\ph}(r_{\ph})=p_{\ph}.
\ee
Expanding the vector potential components in~\eq{momenta} over 
$r-r_\varphi$ up to the linear order 
and using $A^\prime_\vartheta/A^\prime_\varphi=-\rd \psi_{\rm tor}/\rd \psi_{\rm pol}=-q$, 
the poloidal momentum is approximated by
\be{}
p_{\tht}=\frac{e_{\alpha}}{c}A_{\tht}+\frac{m_{\alpha}v_{\parallel}}{h^{\tht}}.
\ee
In the above formula and in the remaining derivation, all quantities
are evaluated at $r=r_{\ph}$ if not noted otherwise. In this approximation
it is possible to express derivatives with respect to $p_{\ph}$ by
radial derivatives. The poloidal action is defined for trapped $(\delta_{\rm t-p}=0)$
and passing $(\delta_{\rm t-p}=1)$ particles by
\be{}
J_{\tht}=\frac{1}{2\pi}\oint d\tht p_{\tht}=\frac{e_{\alpha}}{c}A_{\tht}\delta_{\rm t-p}+J_{\parallel}.
\ee
The first term cancels when integrating back and forth between the
turning points of a trapped orbit. The parallel adiabatic invariant
may be written as a bounce average,
\be{}
J_{\parallel}=\frac{m_{\alpha}\tau_{b}}{2\pi}\left\langle v_{\parallel}^{2}\right\rangle _{b} ,
\ee
with bounce time $\tau_{b}$, orbit time $\tau$ and bounce averaging
$\left\langle a(\tht)\right\rangle _{b}$ defined by
\begin{align}
\tau_{b} & =\oint\frac{dl}{v_{\parallel}}=\oint\frac{d\tht}{v_{\parallel}h^{\tht}},\label{eq:bouncetime}\\
\tau(\tht_{0},\tht_{{\rm orb}}) & =\int_{\vartheta_{0}}^{\tht_{{\rm orb}}}\frac{d\tht}{v_{\parallel}h^{\tht}},\label{eq:orbittime}\\
\left\langle a(\tht)\right\rangle _{b} & =\frac{1}{\tau_{b}}\oint\frac{d\tht}{v_{\parallel}h^{\tht}}a(\vartheta)=\frac{1}{\tau_{b}}\int_{0}^{\tau_{b}}d\tau\, a(\tht_{{\rm orb}}(\tht_{0},\tau)).\label{eq:bounceavg}
\end{align}
Here $a(\tht)$ is any function of the poloidal angle and integrals
of motion $(J_{\perp},H_{0},s_{\ph})$ and $\tht_{{\rm orb}}(\tht_{0},\tau)$
is the (periodic) solution of the unperturbed guiding center equations
(the orbit) starting at the magnetic field minimum point $\tht_{0}$.
Finally, we arrive at the expressions for the three canonical actions
in a tokamak~\cite{Kaufman1972-1063},
\begin{align}
J_{1} & =J_{\perp}=\frac{m_{\alpha}c}{e}\mu,\nonumber \\
J_{2} & =J_{\tht}=\frac{e_{\alpha}}{c}A_{\tht}\delta_{\rm t-p}+\frac{m_{\alpha}\tau_{b}}{2\pi}\left\langle v_{\parallel}^{2}\right\rangle _{b},\nonumber \\
J_{3} & =p_{\ph}=m_{\alpha}v_{\parallel}h_{\ph}+\frac{e_{\alpha}}{c}A_{\ph},
\label{ptor}
\end{align}
where $\mu$ denotes the magnetic moment. Canonical frequencies $\Omega^{k}=\partial H/\partial J_{k}$
are
\be{canfreq}
\Omega^{1} =\left\langle \omega_{c}\right\rangle _{b},
\qquad
\Omega^{2} =\omega_{b},
\qquad
\Omega^{3} =q\omega_{b}\delta_{\rm t-p}+\left\langle v_{g}^{\varphi} \right\rangle_b,
\ee
where the bounce frequency $\omega_{b}=2\pi/\tau_b$ is strictly positive for trapped particles, whereas
for passing particles it can take both, positive and negative values. 
The bounce average of the toroidal precession frequency $v_{g}^{\varphi}$ due to 
the cross-field drift is separated in two parts,
\be{vgphi}
\left\langle v_{g}^{\varphi} \right\rangle_b\equiv \Omega_t = 
\left\langle \frac{v_{\parallel}}{\omega_{c}\sqrt{g}}
\difp{}{r}\left(\frac{v_{\parallel}}{h^{\vartheta}}\right) \right\rangle_b
=\left<\Omega_{tE}\right>_b+\left<\Omega_{tB}\right>_b.
\ee
Here, bounce averages of electric drift frequency $\Omega_{tE}$ and magnetic drift frequency
$\Omega_{tB}$ are 
\begin{align}
\left<\Omega_{tE}\right>_b &= \Omega_{tE} =-\frac{c}{\psi_{{\rm pol}}^{\prime}}\difp{\Phi}{r},\nonumber \\
\left<\Omega_{tB}\right>_b & =\frac{v^{2}}{\psi_{{\rm pol}}^{\prime}}
\left<-\frac{2-\eta B}{2\omega_c}\difp{B}{r}
+\frac{1-\eta B}{\omega_c}h^{\tht}\left(\difp{B_{\tht}}{r}+q\difp{B_{\ph}}{r}+
B_{\ph}\frac{dq}{dr}\right)\right>_b,\label{eq:OmtE}
\end{align}
with equilibrium potential $\Phi$, and velocity space parameterized
by velocity module $v$ and the parameter $\eta=v_{\perp}^{2}/(v^{2}B)=2 e_\alpha J_\perp/(c m_\alpha^2 v^2)$.
Comparison of magnetic rotation frequency $\left<\Omega_{tB}\right>_b$ given by Eq.~\eq{eq:OmtE} with the expression
obtained by bounce averaging of Eq.~(67) of Ref.~\onlinecite{kasilov14-092506} one can notice
the absence in the latter expression of a term $B_{\ph}q^{\prime}(r)$ describing the magnetic shear.
This results from using the local neoclassical ansatz as a starting point
in the linearized equation for the non-Maxwellian perturbation of the distribution function 
where the radial derivative of this perturbation is ignored.
This local ansatz is the standard method in drift kinetic equation solvers in general 3D toroidal
geometries~\cite{hirshman1986-2951,landreman14-042503} and is justified in most transport regimes,
but not in resonant regimes, where magnetic drift plays a significant role.
As shown in the example below, the shear term may lead to a significant modification of the superbanana
resonance condition. This term is retained if linearization is applied after
bounce-averaging the kinetic equation~\cite{shaing15-905810203}.
\red{It should be noted that the guiding center Lagrangian~\eq{gclagr} 
used as a starting point here is valid for the general case of the magnetic field, 
which is not necessarily a force-free field. 
Therefore, the effects of finite plasma pressure on the toroidal rotation 
velocity~\cite{connor83-1702} are automatically taken into account in~\eq{eq:OmtE}.
}

\noindent
The canonical angles in the leading order follow as
\be{eq:angles}
\theta^{1} =\phi-\Delta\phi(\theta^{2},\bJ),
\qquad
\theta^{2} =\Omega^{2}\tau,
\qquad
\theta^{3} =\ph_{H}+q\theta^{2}\delta_{\rm t-p}-q\tht_{{\rm orb}}(\tht_{0},\tau),
\ee
where $\Delta \phi$ is a periodic function of the canonical poloidal variable $\theta^2$.
Since according to~\eq{eq:angles} $\phi$ and $\ph_H$ differ from the respective canonical angles
$\theta^{1}$ and $\theta^{3}$ by additional terms depending on $\theta^2$ only
and $\tht$ depends only on $\theta^{2}$,
the spectrum $a_{\mathbf{m}}$ in canonical angles
of a function given by a single harmonic (l,n) of the original angles $\phi,\ph$,
\be{spectrum}
a(\phi,\tht,\ph)=a_{ln}(\vartheta)e^{i\left(l\phi+n\ph\right)}=\sum_{\mathbf{m}}a_{\mathbf{m}}e^{im_{k}\theta^{k}},
\ee
contains non-zero contributions only from canonical modes with $m_{1}=l$ and $m_{3}=n$.
In particular, for the gyroaverage $\left\langle a\right\rangle _{g}$
described by the harmonic $l=0$ of function $a$,
one obtains to the leading order in $\rho_{\parallel}$
\be{eq:amn}
a_\bm=\left\langle a_{0n}(\tht)e^{inq\tht-i\left(m_{2}+nq\delta_{\rm t-p}\right)\omega_{b}\tau}\right\rangle _{b},
\ee
where $\bm=(0,m_{2},n)$.

\section{Neoclassical toroidal viscous torque and related radial transport}
\label{sec:neotorque}

\noindent
For NTV applications, where the perturbed Hamiltonian~\eq{H0deltaH} is independent of gyrophase,
only harmonics with first canonical mode number $m_{1}=0$ can contribute
to fulfill the resonance condition inside the $\delta$ distribution
of Eq. (\ref{eq:Qm}), and the latter is reduced to
\be{eq:resonance}
m_{j}\Omega^{j}=0\,\,\,\rightarrow\,\,\, (m_{2}+nq\delta_{\rm t-p})\omega_{b}+n\Omega_t=0.
\ee
This equation includes all regimes of interest here: The superbanana-plateau
resonance is described by the condition $m_{2}=0$ for trapped particles.
For passing particles, $m_{2}=0$ corresponds to a transit resonance.
This is the only resonance remaining in the infinite aspect ratio
limit, where it reduces to the usual Cherenkov (TTMP) resonance. Finite
mode numbers $m_{2}$ correspond to bounce and bounce-transit resonances
for trapped and passing particles, respectively. Resonances where both, 
parallel motion and cross-field drift determine the resonance condition,
i.e. all resonances except the superbanana-plateau resonance
are mentioned below as ``drift-orbit'' resonances.

\noindent
Due to the properties of the spectrum~\eq{spectrum}, which follow from the axial symmetry
of the unperturbed field, separate toroidal 
harmonics of perturbation Hamiltonian produce independent contributions
to the torque and particle flux density. Therefore it is sufficient
to assume the perturbation field $\delta B$ in~\eq{H0deltaH} in the form
of a single toroidal harmonic,
\be{}
\delta B={\rm Re}(B_{n}(\tht)e^{in\ph}).
\ee
Making use of Eq. (\ref{eq:amn}), the associated modes
of the Hamiltonian perturbation result in
\be{eq:Hm}
H_{\mathbf{m}}=\left\langle \left(m_{\alpha}v_{0\parallel}^{2}(\tht)
+\frac{e_{\alpha}}{m_{\alpha}c}J_{\perp}B_{0}(\tht)\right)\frac{B_{n}(\tht)}{B_{0}(\tht)}
e^{inq\tht-i\left(m_{2}+nq\delta_{\rm t-p}\right)\omega_{b}\tau}\right\rangle _{b}.
\ee
For small enough perturbations, which are considered here, quasilinear
effects are weak and thus $f_{0}$ is close to a drifting Maxwellian,
\be{f0def}
f_{0}=\frac{n_{\alpha}}{\left(2\pi
m_{\alpha}T_{\alpha}\right)^{3/2}}e^{(e_{\alpha}\Phi-H_{0})/T_{\alpha}},
\ee
with parameters depending on $r_\varphi$ but not $r$. This Maxwellian differs
from a local Maxwellian by linear terms in $\rho_\parallel$, which, as shown below,
provide negligible contributions in resonant regimes with quasi-static perturbations.
\red{
Let us check that within first order in $\rho_\parallel$ 
the toroidally drifting Maxwellian
Eq.~\eq{f0def} 
is a solution to the axisymmetric kinetic 
equation for ions valid in all collisionality regimes in the absence of temperature
gradients.
For this purpose it is more convenient to replace
the approximate expression for the canonical angular momentum~\eq{ptor} valid in
first order in Larmor radius by the exact expression,
\be{ptorex}
p_\varphi=m_\alpha v_\varphi +\frac{e_\alpha}{c}A_\varphi,
\ee
where $v_\varphi$ is the toroidal covariant component of the total particle velocity
including the Larmor gyration and $A_\varphi$ is evaluated at the exact particle position
$r_{\rm c}$ (not at the guiding center position denoted with $r$ here)
related to $r_\varphi$ as follows,
\be{exradtorphi}
r_{\rm c}=r_\varphi+\frac{c m_\alpha v_\varphi}{e_\alpha \sqrt{g}B^\vartheta}.
\ee
Then, 
the unperturbed distribution function~\eq{f0def} 
up to linear order in Larmor radius is
\be{f0uptolin}
f_{0}=\frac{n_{\alpha}}{\left(2\pi
m_{\alpha}T_{\alpha}\right)^{3/2}}
\exp\left(-\frac{m_\alpha}{2T_\alpha}
\left(\bv_{\rm pol}^2+g_{\varphi\varphi} \left(v^\varphi-V_\alpha^\varphi\right)^2\right)
\right),
\ee
where $g_{\varphi\varphi}=R^2$ and all functions of radius are evaluated at $r_c$.
Here, $\bv_{\rm pol}$ and $v^\varphi$ are total poloidal and contra-variant component
of the total toroidal particle velocity, respectively, and the contra-variant toroidal 
component of the ion flow velocity is explicitly given by
\be{torionflow}
V_\alpha^\varphi = \frac{c}{\sqrt{g}B^\vartheta}
\left(E_r-\frac{T_\alpha}{e_\alpha n_\alpha}\difp{n_\alpha}{r_c}\right).
\ee
In a simple plasma where the momentum is approximately conserved within a single
ion component, the drifting Maxwellian~\eq{f0uptolin} annihilates the collision term. 
As a straightforward consequence of~\eq{f0uptolin}, the poloidal ion flow velocity
is zero at all collisionalities if the temperature gradient is absent. Respectively,
the toroidal flow velocity~\eq{torionflow} is the same as given by ideal MHD
(see, e.g., Eq.~(6) of Ref.~\onlinecite{kasilov14-092506}).
In the presence of temperature gradients and in a multi-species plasma Eq.~\eq{f0def}
satisfies the kinetic equation only in zero order over Larmor radius. Additional
anisotropic terms which appear in the first order solution are of same order as 
in~\eq{f0def}, nevertheless, they provide a negligible contribution for the following reason.
}
When substituting~\eq{f0def} in~\eq{eq:Qm} one can notice that only derivatives
of the parameters over $r_\varphi$ provide non-zero contributions in presence
of resonance condition,
\be{substres}
\delta(m_j\Omega^j) m_{k}\difp{f_{0}}{J_{k}}=
- \delta(m_j\Omega^j) \frac{nc(A_{1}+A_{2}u^{2})}{e_\alpha}
\frac{\rd r}{\rd \psi_{\rm pol}}f_0,
\ee
where $u=v/v_{T}$ is the velocity module $v$ normalized by the thermal velocity $v_{T}=\sqrt{2T_\alpha/m_\alpha}$ and
\be{thermforces}
A_{1}=\frac{1}{n_{\alpha}}
\difp{n_{\alpha}}{r}+\frac{e_{\alpha}}{T_{\alpha}}\difp{\Phi}{r}-
\frac{3}{2T_{\alpha}}\difp{T_{\alpha}}{r},
\qquad
A_{2}=\frac{1}{T_{\alpha}}\difp{T_{\alpha}}{r},
\ee
are the thermodynamic forces which are evaluated at $r=r_\varphi$.
For any function $F$ of actions expressed in the form $F=F(H_0,J_\perp,p_\varphi)$, 
only the derivative over $p_\varphi$ remains in expressions such as~\eq{substres} because
the derivative over $J_1$ enters with factor $m_1=0$ only, and the derivative over $H_0$ enters with
factor $m_k\Omega^k$ which is zero due to the resonance condition~\eq{eq:resonance}
(the energy is preserved for static perturbations). Therefore the contribution of the linear 
correction in $\rho_\parallel$ to the unperturbed distribution function which depends also 
on $J_\perp$ would contribute in~\eq{substres} only in the form of its derivative over $r_\varphi$
which is of higher order in $\rho_\parallel$ than such a derivative of the Maxwellian retained
in~\eq{substres}.
In the expression for the torque density~\eq{torque_ql} one can ignore finite Larmor radius
effects together with finite orbit width effects in $r_c$ in the argument of the $\delta$-function
by setting $r_c \approx r_\varphi$. Then an integration over $J_3=p_\varphi$ results in a replacement
of $r_\varphi$ by $r$ in the subintegrand, and the integration over canonical angles is simply replaced
by a factor $8\pi^3$. Changing the integration variables of the remaining integral over $J_1$ and $J_2$ 
to $v$ and $\eta$ and transforming the resulting $T_\ph^{\rm NA}$ to a particle flux density using the flux-force 
relation~\eq{eq:FluxForce} results in
\be{Gamma_ntv}
\Gamma = \frac{2\pi^2 n m_{\alpha}^3 c}{e_\alpha S}
\int_{0}^{\infty}dv\, v^{3}\int_{0}^{1/B_{{\rm min}}}d\eta\tau_{b}\sum_{{\rm m}_{2}}Q_{\mathbf{m}}.
\ee
Substituting $Q_{\mathbf{m}}$ in~\eq{Gamma_ntv} explicitly
and using the representation of $\Gamma$ in terms of thermodynamic forces~\eq{thermforces},
$\Gamma=-n_\alpha(D_{11}A_{1}+D_{12}A_{2})$, resonant transport coefficients follow as
\be{eq:D1x}
D_{1k}=\frac{\pi^{3/2}n^{2}c^{2}v_{T}}{e_{\alpha}^{2}S}\frac{\rd r}{\rd \psi_{\rm pol}}
\int_{0}^{\infty}du\, u^{3}e^{-u^{2}}
\sum_{m_{2}}\sum_{{\rm res}}\left(\tau_{b}|H_{\mathbf{m}}|^{2}
\left|m_{2}\frac{\partial\omega_{b}}{\partial\eta}+n\difp{\Omega^{3}}{\eta}\right|^{-1}\right)_{\eta=\eta_{{\rm res}}}w_{k},
\ee
where $w_{1}=1$ for $D_{11}$ and $w_{2}=u^{2}$ for $D_{12}$, respectively.
In this expression the $\delta$ term inside $Q_{\mathbf{m}}$ has
been evaluated with respect to $\eta$, and $\eta_{\mathrm{res}}$ are
(generally multiple) roots of Eq. (\ref{eq:resonance}).

\noindent
In the direct definition of the flux~\eq{partlux_ql} one can, again, replace $r_c$ by $r_\varphi$
in the argument of the $\delta$-function. Using the same arguments as in~\eq{substres}
for ignoring the linear order term in $\rho_\parallel$ inside $f_0$, one can ignore the difference between
$r_c$ and $r_\varphi$ in the derivative $m_k \partial r_c /\partial J_k$. Then $\Gamma_{\rm F}$ given 
by~\eq{partlux_ql} leads to a result identical to~\eq{Gamma_ntv}.

\red{
\noindent
The equivalence of $\Gamma_F$ and $\Gamma$ obtained here using a simple Krook collision model
indicates that momentum conservation plays no role in resonant transport.
While in case of superbanana plateau and bounce resonance regimes 
this can be concluded, in particular, from
the fact that all resonant particles are trapped particles, which lose 
parallel momentum obtained from the perturbation field within a single bounce period
due to magnetic mirroring, this explanation cannot be used for transit and bounce-transit
resonances where passing particles are responsible. The general reason for the 
conclusion above
is different and is actually the same as the reason to ignore the anisotropic correction
in the unperturbed distribution function~\eq{f0def}. As already mentioned, the resonant
interaction with a static perturbation field does not modify the total particle energy $H_0$
in contrast to the case of time dependent perturbations where
the change of total energy scales with perturbation frequency $\omega$
due to the more general resonance condition $m_j\Omega^j=\omega$.
In addition, since cyclotron resonances, $m_1\ne 0$, cannot be realized for bulk 
particles (particles with energies of the order of thermal energy), the perpendicular
adiabatic invariant $J_\perp$ is also conserved. Consequently, the change of parallel
velocity $v_\parallel=v_\parallel(\bx,H_0,J_\perp)$ and of bounce frequency
$\omega_b=\omega_b(r_\varphi,H_0,J_\perp)$, which represents the bounce averaged parallel
momentum of passing particles, can only appear through the change
by the resonant interaction of the particle position in space $\bx$ (``radial'' variable $r_\varphi)$. 
In case of mild radial electric fields
where the variation of the potential energy along the guiding center orbit with finite radial width
is small (of the order of Larmor radius) compared to the thermal energy what corresponds
to sub-sonic rotations (small toroidal Mach numbers), 
the contribution of the kinematic momentum change to the overall canonical
momentum change is small of the same order too. Therefore, the momentum restoring term
in the collision operator provides a correction proportional to the toroidal Mach number
assumed to be small in the present paper.
}

\noindent
As mentioned above, the Hamiltonian approach includes all quasilinear resonant transport regimes in 
a unified form where these regimes correspond to different resonances~\eq{eq:resonance}.
In particular the expression for the contribution of the $m_2=0$ resonance for trapped particles
corresponds to the superbanana-plateau regime and differs from such a result of 
Ref.~\onlinecite{shaing15-905810203} only in notation. The results for drift-orbit resonances 
$m_2\ne 0$ mostly agree with Ref.~\onlinecite{shaing09-075015} 
up to simplifications of the magnetic field geometry and the neglected magnetic drift 
in this reference. 
Differences appear only in resonant contribution of passing particles
on irrational flux surfaces arising from the representation in Eqs.~(37) and~(38) of 
Ref.~\onlinecite{shaing09-075015} of an aperiodic function by a Fourier series.

\section{Numerical implementation and results}
\label{sec:numerics}

\noindent
In the scope of this work the coefficients (\ref{eq:D1x}) are computed numerically 
in the newly developed code NEO-RT for the general case of a perturbed tokamak 
magnetic field specified in Boozer coordinates.
Bounce averages are performed via numerical time integration of zero order
guiding center orbits as specified in~\eq{eq:amn}. An efficient
numerical procedure for finding the roots in Eq. (\ref{eq:resonance})
is realized using the scalings
\bea{}
\omega_{b} & =u\bar{\omega}_{b}(\eta),\\
\left<\Omega_{tB}\right>_b & =u^{2}\bar{\Omega}_{tB}(\eta).
\eea
Normalized frequencies $\bar{\omega}_{b}$ and $\bar{\Omega}_{tB}$
(relatively smooth functions) are precomputed on an adaptive $\eta$-grid
and interpolated via cubic splines in later calculations.

\begin{figure}
\includegraphics{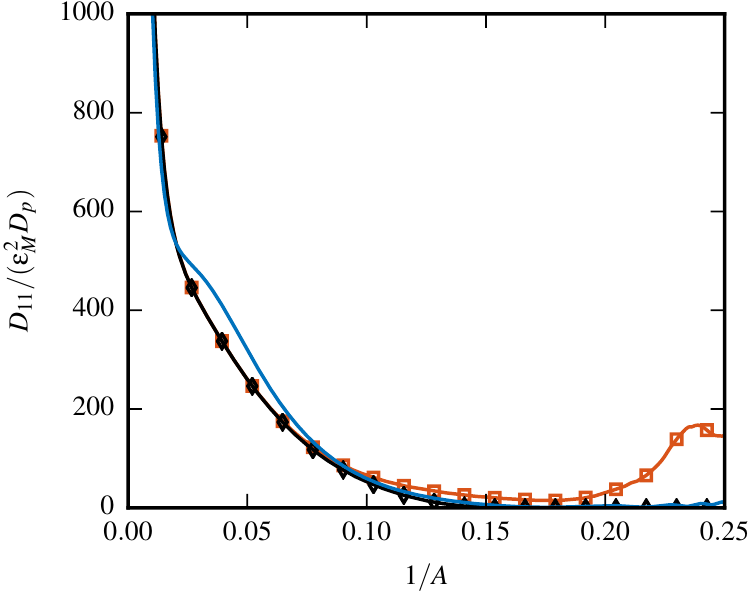}\quad{}\includegraphics{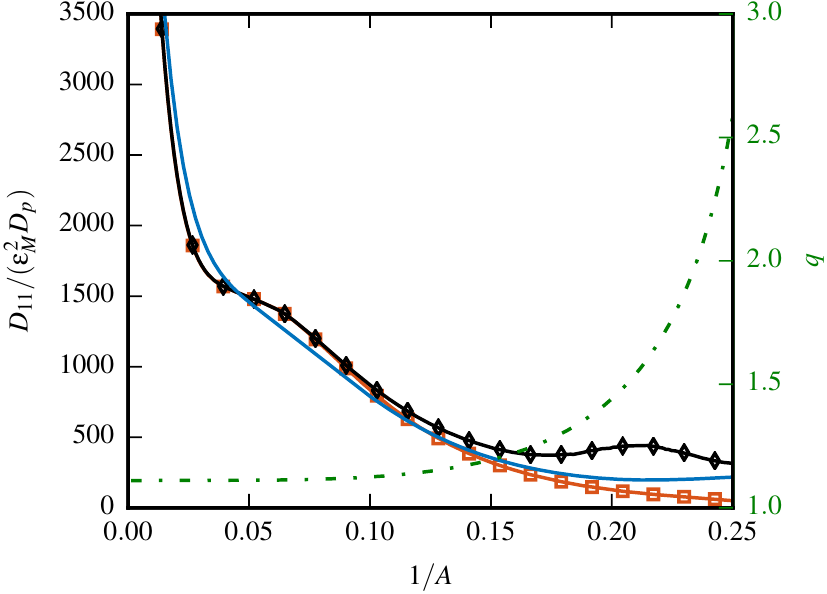}\\
\includegraphics{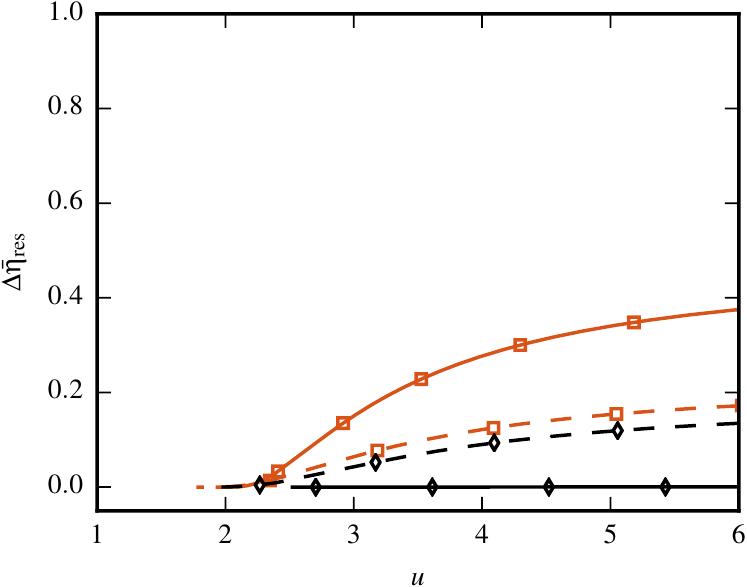}\quad{}\includegraphics{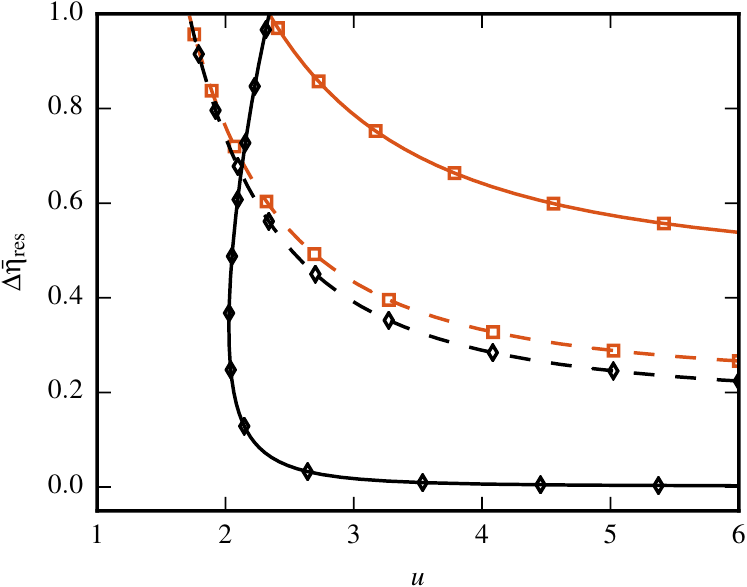}

\protect\caption{Radial dependence of superbanana plateau $D_{11}$ in the 
RMP case for Mach number $M_t=0.036$ (left) and $-0.036$ (right). Comparison of Hamiltonian approach (NEO-RT)
to analytical formula by Shaing~\cite{shaing09-035009} (solid line). Results with ($\diamond$) and without 
magnetic shear ($\square$) in the magnetic drift frequency (\ref{eq:OmtE}). A 
safety factor profile (dash-dotted) is shown on the second axis of the upper right plot. 
The lower plots show resonance lines ranging from deeply trapped ($\Delta\bar{\eta}=0$)
to trapped passing boundary ($\Delta\bar{\eta}=1$) at  
flux surfaces of aspect ratio $A=5$ (solid) and $A=10$ (dashed).  \label{fig:sb-plateau}}
\end{figure}

\noindent
For testing and benchmarking, a tokamak configuration 
with circular concentric flux surfaces and safety factor shown in 
Fig.~\ref{fig:sb-plateau} is used (the same as in Ref.~\onlinecite{kasilov14-092506})
and results are compared to calculations from the NEO-2 code. 
The perturbation field amplitude in Eq. (\ref{eq:Hm}) is
taken in the form of Boozer harmonics
\be{}
B_{n}(\tht)=\varepsilon_{M}B_{0}(\tht)e^{im\tht}.
\ee
Two kinds of perturbations are considered here: a large scale perturbation
with $(m,n)=(0,3)$ referred below as ``RMP-like case'' because of the toroidal
wavenumber typical for perturbations produced by ELM mitigation coils,
and a short scale perturbation with $(m,n)=(0,18)$ typical for the 
toroidal field (TF) ripple.
The remaining parameters are chosen to be representative for a realistic medium-sized tokamak configuration.
In the plots, transport coefficients $D_{1k}$ are normalized by (formally infinitesimal)
 $\varepsilon_{M}^{2}$ times the mono-energetic plateau value 
\be{plateau}
D_{p}=\frac{\pi qv_{T}^{3}}{16\, R\,\bar{\omega}_{c}^{2}},
\ee
where $R$ is the major radius, and the reference gyrofrequency $\bar{\omega}_{c}$
is given by the $(0,0)$ harmonic of $\omega_{c}$. 
Radial dependencies are represented by the flux surface aspect ratio 
$A=\left(\psi^a_{\rm tor}/\psi_{\rm tor}\right)^{1/2}R/a$ 
of the current flux surface where $a$ is the minor radius of the outermost flux surface
and $\psi^a_{\rm tor}$ the toroidal magnetic flux at this surface. 
The radial electric field magnitude is given in terms of
the toroidal Mach number $M_t \equiv R \Omega_{tE}/v_T$. 
In all plots there are at least 4 data points between subsequent markers. 

\noindent
Fig.~\ref{fig:sb-plateau} shows the radial dependence of the transport coefficient 
$D_{11}$ in the superbanana plateau regime for the RMP-like 
perturbation for both positive and negative radial electric field.
For this benchmarking case the relation between toroidal precession frequencies
due to the $\bE\times\bB$ drift, $\Omega_{tE}$, and due to the magnetic drift $\Omega_{tB}$,
has been fixed by setting the reference toroidal magnetic drift frequency $\Omega^{\rm ref}_{tB}\equiv cT_{\alpha}/(e_{\alpha} \psi^a_{\rm tor})$ (not the actual $\Omega_{tB}$) equal to $\Omega_{tE}$.
Additional curves are shown for calculations where the
 magnetic shear term ($dq/dr$) in Eq.~\eq{eq:OmtE} has been neglected. The results 
are compared to the analytical formula for the large aspect ratio limit  by Shaing~\cite{shaing09-035009}.
Resonance lines in velocity space are plotted below the radial profiles for a flux surface
relatively close to the axis ($A=10$) and one further outwards ($A=5$). 
Here $\Delta\bar{\eta}=(\eta-\eta_{\mathrm{tp}})/(\eta_{\mathrm{dt}}-\eta_{\mathrm{tp}})$ is the 
distance to the trapped passing boundary $\eta_{\mathrm{tp}}$ normalized to the trapped 
region between trapped-passing boundary $\eta_{\mathrm{tp}}$ and deeply trapped $\eta_{\mathrm{dt}}$.
For flux surfaces with $A>10$ 
magnetic shear plays a small role due
to the flat safety factor profile in the present field configuration:
The diffusion coefficient $D_{11}$ is nearly identical to the result without shear 
and stays close  to the analytical result for the large aspect ratio limit. 
For aspect ratio $A=10$,
the agreement between NEO-2 calculations and large aspect ratio limit of Ref.~\onlinecite{shaing09-035009}
has been demonstrated earlier in Ref.~\onlinecite{kasilov14-092506}.
At larger radii, where the q profile becomes steep, a significant deviation between
the cases with and without magnetic shear term is visible.
This can be explained by the strong shift of the resonance lines due to the shear term
in the rotation frequency $\Omega_{tB}$ that is visible in lower plots. For both signs of the 
electric field, the resonant $\eta_{\rm res}$ is closer to the trapped passing boundary 
when shear is included. 

\begin{figure}
\includegraphics{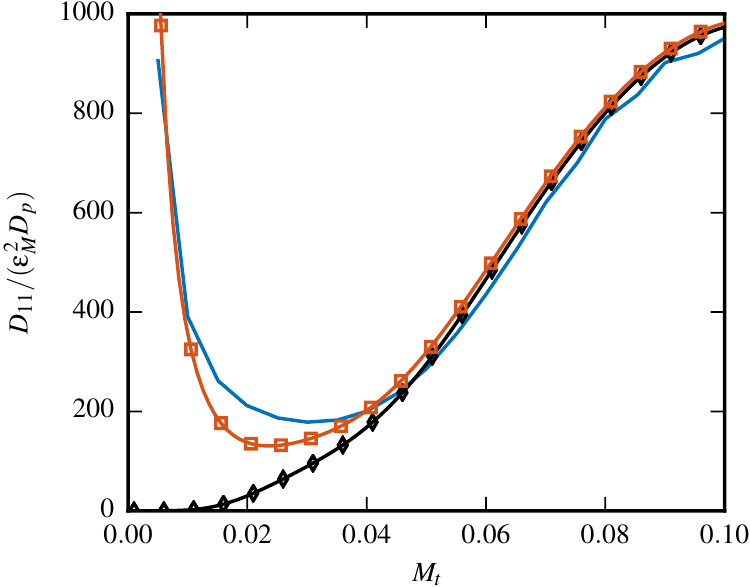}\quad{}\includegraphics{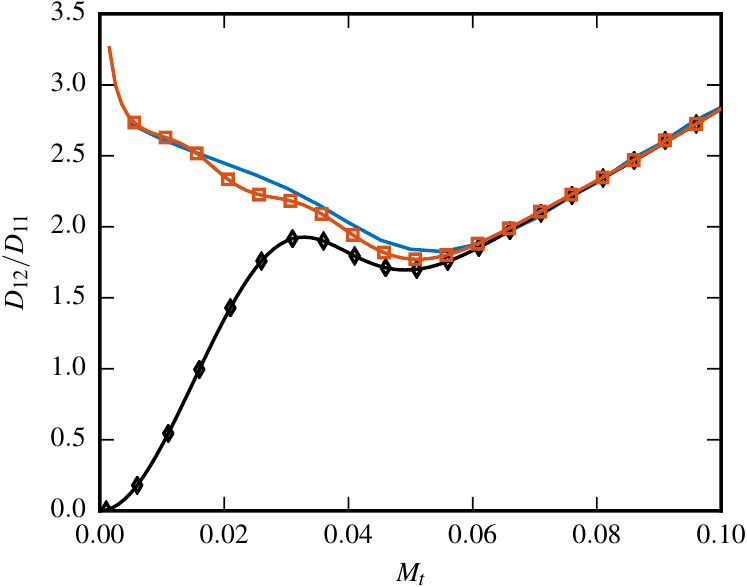}

\protect\caption{Drift-orbit resonances with neglected magnetic drift:
Mach number dependence of $D_{11}$ (left) and the ratio $D_{12}/D_{11}$ (right) 
for an RMP-like perturbation at $A=10$. 
Comparison of Hamiltonian approach ($\diamond$), sum of
Hamiltonian results and $\nu-\sqrt{\nu}$ regime by Shaing~\cite{shaing10-025022} ($\square$),
and results from NEO-2 at collisionality $\nu^\star=3\cdot10^{-4}$ (solid line).
\label{fig:do_transport}}
\end{figure}

\begin{figure}
\includegraphics{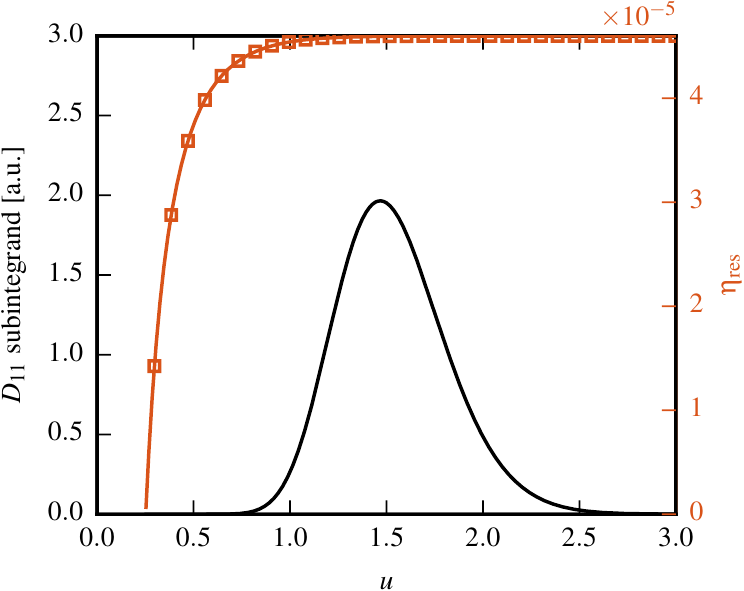}\quad{}\includegraphics{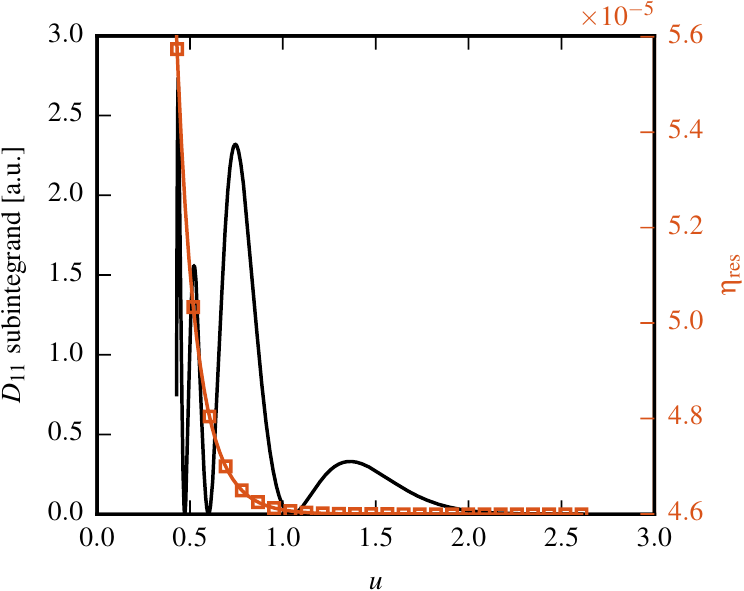}

\protect\caption{Drift-orbit resonances, RMP at $A=10$ with $M_t=0.028$.
Dependence of the subintegrands in Eq.~\eq{eq:D1x} on the normalized velocity $u$
for the dominant mode (solid line) of passing (left, $m_{2}=-3$) 
and trapped particles (right, $m_{2}=-1$) and resonance lines for these modes ($\square$,~right axis).
Significant contributions are visible where the resonance is close to the trapped passing boundary
$\eta_{\rm tp} = 4.6\cdot 10^{-5}$.
\label{fig:do_integral}}
\end{figure}

\noindent
In Figs.~\ref{fig:do_transport}-\ref{fig:do_integral} the radial electric field dependence
 of non-ambipolar transport induced by drift-orbit resonances with magnetic drift neglected
($\Omega_{tB}$ set to zero) is pictured. Here, several canonical modes $m_2$ contribute for both, trapped 
and passing particles.

\noindent
In Fig.~\ref{fig:do_transport} the Mach number dependence of transport 
coefficient $D_{11}$ and the ratio $D_{12}/D_{11}$ is plotted for this regime
for an RMP-like perturbation ($n=3$). NEO-2 calculations shown for the comparison
have been performed at rather low collisionality (see the caption) characterized by
the parameter $\nu^\ast=2\nu q R/v_T$ where $\nu$ is the collision frequency.
In addition, also the curves with the sum of diffusion coefficients in the 
collisional $\nu-\sqrt{\nu}$ regime from the joint formula of Shaing~\cite{shaing10-025022}
and resonant contributions from the Hamiltonian approach are shown.
For $M_t<0.02$, in contrast to the superbanana plateau regime, collisionless 
transport is small compared to collisional effects.
Between $M_t=0.02$ and $0.04$ the sum of Hamiltonian and $\nu-\sqrt{\nu}$
results for $D_{11}$ is clearly below NEO-2 values. The reason for this are contributions
near the trapped passing boundary, which are illustrated in Fig.~\ref{fig:do_integral} 
at $M_t=0.028$. There the integrand in Eq.~\eq{eq:D1x} for the mode $m_2$ with the 
strongest contribution is shown together with the resonance line in velocity space. 
For $M_t>0.04$ there is a close match between the results with slightly lower $D_{11}$ 
values from NEO-2 due to remaining collisionality effects.

\red{
\noindent
It should be noted that validity of the ``collisionless'' Hamiltonian model
cannot be accessed with the help of a simple Krook model although this model is fully
adequate for the present derivations. The details of the collision model are not important
as long as the collisional width of the resonant line in velocity space is 
smaller than the distance from that line to the trapped-passing boundary where
the topology of the orbits changes abruptly. This criterion is much more restrictive
than the smallness of the collision frequency compared to the bounce frequency suggested
by the Krook model. At small Mach numbers where the resonant line approaches
the trapped-passing boundary rather closely (see Fig.~\ref{fig:do_integral}), the applicability 
of the ``collisionless'' approach
is violated at much lower collisionalities than one could expect from the Krook model,
and in that case a collisional boundary layer analysis including the resonant interaction 
is needed. As one can see from Fig.~\ref{fig:do_transport}, for such transitional Mach 
numbers where both, $\nu-\sqrt{\nu}$ regime and resonant regime are important, a simple
summation of the separate contributions from these regimes obtained in asymptotical limits
cannot reproduce the numerical result, similarly to the observation in
Ref.~\onlinecite{sun10-145002}. With increasing Mach numbers, the resonant curve gets more
separated from the trapped-passing boundary, and the collisionless analysis becomes
sufficient, as it can be seen for higher Mach numbers in Fig.~\ref{fig:do_transport}.
}

\begin{figure}
\includegraphics{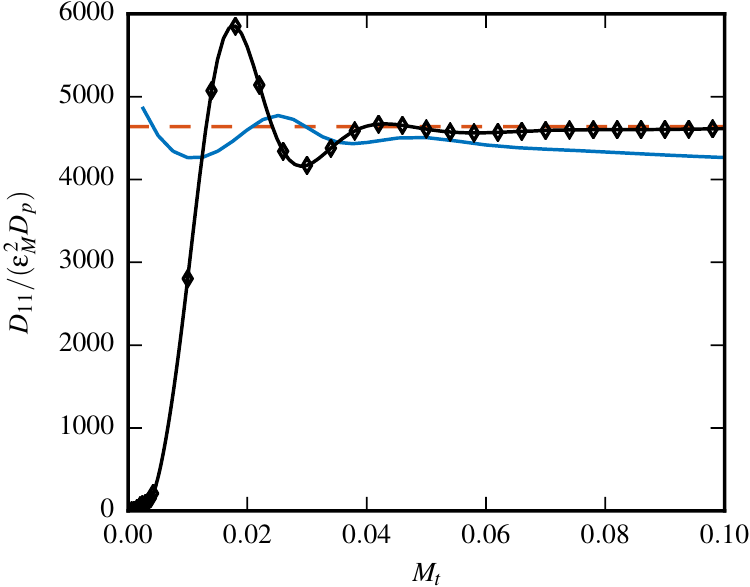}\quad{}\includegraphics{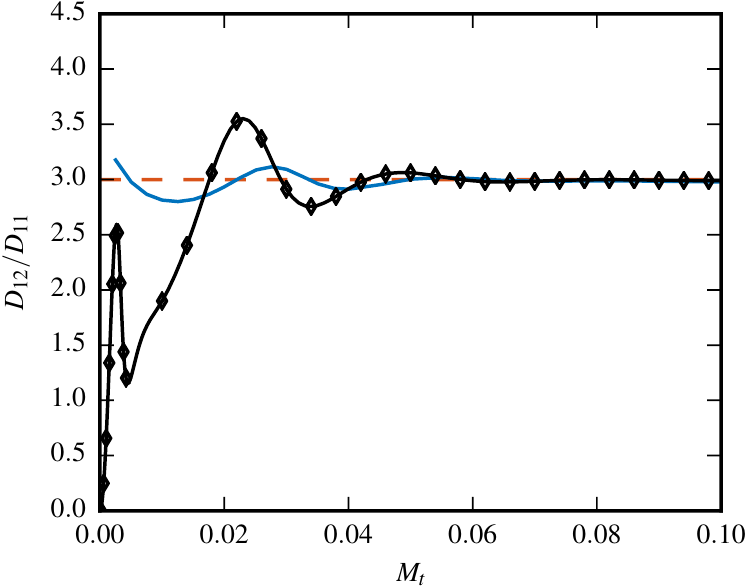}

\protect\caption{Mach number dependence of transport coefficients of drift-orbit resonances
for a toroidal field ripple at $A=10$. Comparison between Hamiltonian approach ($\diamond$),
ripple plateau (dashed) and NEO-2 at collisionality $\nu^\star=10^{-3}$ (solid line).
\label{fig:do_transport18}}
\end{figure}

\noindent
Fig.~\ref{fig:do_transport18} shows the Mach number dependence of $D_{11}$ as well as
$D_{12}/D_{11}$ for a toroidal field ripple ($n=18$) together with the analytical
ripple plateau value~\cite{Boozer1980-2283} and results for finite collisionality from NEO-2.
At low Mach numbers $M_t<0.01$ collisional effects are again dominant. A resonance peak of passing
particles is visible for $D_{12}/D_{11}$ at $M_t=2.8\cdot10^{-3}$. In the intermediate
region between $M_t=0.01$ and $0.05$ oscillations due to trapped particle resonances
are shifted and reduced in the collisional case. For $M_t>0.05$ Hamiltonian results
converge towards the ripple plateau. A small deviation of NEO-2 values for $D_{11}$, which is 
of the order of Mach number is caused by the low Mach number approximation used in NEO-2.

\begin{figure}
\includegraphics{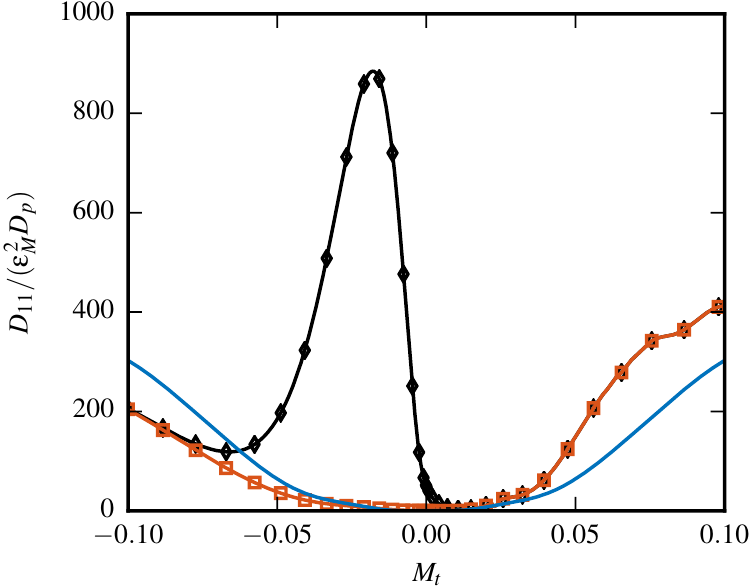}\quad{}\includegraphics{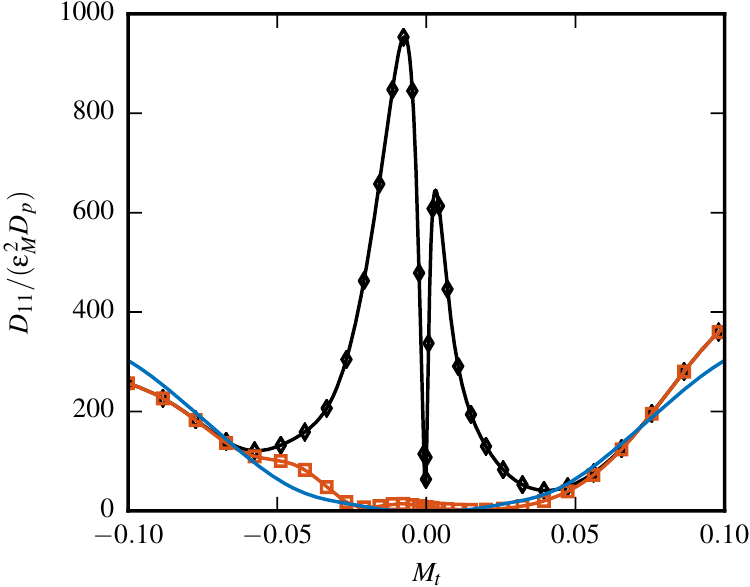}

\protect\caption{Mach number dependence of $D_{11}$ for RMP at $A=5$ with shear term included (left)
and neglected (right) in Eq.~\eq{eq:OmtE}. Total resonant transport ($\diamond$) and 
contributions by drift-orbit resonances with finite magnetic drift and 
excluding superbanana plateau ($\square$).
Comparison to drift-orbit resonances with $\Omega_{tB}$ set to zero (solid line).
\label{fig:n3_A5}}
\end{figure}

\noindent
Finally, in Fig.~\ref{fig:n3_A5} the Mach number dependence of $D_{11}$ for the RMP case
is plotted for both, positive and negative Mach numbers for finite toroidal precession 
frequency due to the magnetic drift $\Omega_{tB}$. To set the scaling with respect
to $\Omega_{tE}$, the reference magnetic drift frequency
defined above is fixed by
$R\Omega^{\rm ref}_{tB}/v_T = 3.6\cdot10^{-2}$.
In this case all resonance types contribute to transport coefficients. 
Due to the finite magnetic drift, the Mach number dependence is not symmetric anymore.
If shear is neglected in  Eq.~\eq{eq:OmtE}, the superbanana plateau 
is centered around slightly negative values of the electric field,
and magnetic drift induces some deviation 
from the idealized case without magnetic drift. In the case with included shear
superbanana plateau, contributions for positive Mach numbers vanish and a large
deviation from the case without magnetic drift is visible also for drift-orbit resonances.

\section{Conclusion}
\label{sec:conclosion}

\noindent
In this article, a method for the calculation of the toroidal torque in low-collisional
resonant transport regimes due to non-axisymmetric perturbations in
tokamaks based on a quasilinear Hamiltonian approach has been presented.
This approach leads to a unified description of all those regimes 
including superbanana plateau and drift-orbit resonances without 
simplifications of the device geometry. Magnetic drift effects including 
non-local magnetic shear contributions are consistently taken into account. 
An efficient numerical treatment is possible by pre-computation 
of frequencies appearing in the resonance condition.

\noindent
The analytical expressions for the transport coefficients obtained within the Hamiltonian
formalism agree with the corresponding expressions obtained earlier for particular resonant
regimes within the validity domains of those results. In particular, the agreement with formulas 
for the superbanana plateau regime, which have been updated recently for a general
tokamak geometry in Ref.~\onlinecite{shaing15-905810203}, is exact.
Minor inconsistencies in the treatment of passing particles 
have been found (see section~\ref{sec:neotorque}) in comparison to the analytical formulas 
for bounce-transit resonances of Ref.~\onlinecite{shaing09-075015}.
\red{
In addition, it has been demonstrated that momentum conservation of the collision operator
plays a minor role in resonant regimes in general as long as the toroidal rotation is
sub-sonic.
}

\noindent
Results from the newly developed code NEO-RT based on the presented 
Hamiltonian approach agree well with the results from the NEO-2 code
at relatively high Mach numbers where finite collisionality effects are small ($M_t>0.04$
in the examples here). At these Mach numbers, both approaches also reproduce the analytical 
result for the ripple plateau regime~\cite{Boozer1980-2283} well.
At intermediate Mach numbers $0.02<M_t<0.04$ which correspond to the transition between
the $\nu-\sqrt{\nu}$ regime and resonant diffusion regime, the combined torque 
of $\nu-\sqrt{\nu}$ regime and resonant diffusion regime does not 
reach the numerical values calculated by NEO-2 even at very low 
collisionalities due to the contribution of the resonant phase space region very 
close to the trapped-passing boundary. Collisional boundary layer analysis is required
in addition to obtain more accurate results in these regions.

\noindent
Within the Hamiltonian approach, which is non-local by its nature, i.e. it does not
use truncated ``local'' orbits which stay on magnetic flux surfaces,
an additional term describing the influence of magnetic shear that is absent in 
the standard local neoclassical ansatz naturally arises in the resonance condition. 
This term significantly increases the asymmetry of the superbanana plateau resonance with 
respect to the toroidal Mach numbers of $\bE\times\bB$ rotation and may even eliminate
this resonance for a given Mach number sign (at positive Mach numbers in the examples here).
This shear term has been included into analytical treatment recently~\cite{shaing15-905810203}
but was absent in earlier approximate formulas~\cite{shaing09-035009,shaing10-025022}.
This could be a possible reason for the discrepancy with the non-local $\delta f$ Monte Carlo 
approach observed in Ref.~\onlinecite{satake11-055001}.

\red{
\noindent
It should be noted that term ``non-local transport ansatz'' is used here with respect 
to the orbits employed in the computation of the perturbation of the distribution function, 
and it should not be confused with the nonlocal transport in the case where the orbit width
is comparable to the radial scale of the parameter profiles and where the transport equations
cannot be reduced to partial differential equations. In the sense used here, the shear term
appears due to a radial displacement of the guiding center, what is a nonlocal effect. Namely,
due to variation of the safety factor with radius, the toroidal connection length between the
banana tips of the trapped particle is different at the outer and the inner sides of the flux 
surface containing these tips. Since particles with positive and negative 
parallel (and, respectively, toroidal) velocity signs are displaced from this surface 
in different directions, the sum of the toroidal displacements over the full banana orbit 
is not balanced to zero, what results in an overall toroidal drift proportional to the
shear parameter.
This effect cannot be described by the local ansatz in an arbitrary coordinate system
but still can be retained within the local ansatz in the field aligned coordinates
as the ones used in Ref.~\onlinecite{shaing15-905810203}. This ambiguity in 
the description of the magnetic drift within the local ansatz~\cite{hokan} results from the fact
that setting to zero one of the velocity vector components which are not invariant under a  
coordinate transformation destroys the covariance of equations of motion during such transformations.
}

\noindent
Magnetic shear can also have a strong influence on drift-orbit (bounce and bounce-transit)
resonances. A comparison between the results in this regime with neglected magnetic drift and 
results including magnetic drift shows a strong discrepancy, 
especially if magnetic shear is considered.  Therefore, for an accurate evaluation of
NTV torque in low-collisional resonant transport regimes it is necessary to consider
magnetic drift including magnetic shear in the resonance condition. This is especially 
important for modern tokamaks with poloidal divertors where magnetic shear is high at 
the plasma edge where the main part of the NTV torque is produced.

\noindent
It should be noted that benchmarking with NEO-2 performed in this work resulted in
improvement of the analytical quasilinear approach~\cite{kasilov14-092506} used in NEO-2 as 
well as in the numerical treatment. In particular, the use of compactly supported basis functions~\cite{kernbichler-ppcf}
for the discretization of the energy dependence of the distribution function instead of global Laguerre 
polynomials used in earlier
NEO-2 versions allowed to obtain correct results also at high Mach numbers with rather low collisionality
where the global basis resulted in artificial oscillations of the diffusion coefficients with Mach number.
In addition, the standard local neoclassical approach used in Ref.~\onlinecite{kasilov14-092506} for 
the derivation of quasilinear equations has been generalized to a non-local approach where the effect 
of magnetic shear is treated appropriately. Details of the derivation will be published in a 
separate paper. NEO-2 results for NTV in an ASDEX Upgrade equilibrium from both, local and 
non-local approach are shown and compared in Ref.~\red{\onlinecite{martitsch16-074007}}.

\begin{acknowledgments}
This work has been carried out within the framework of the EUROfusion Consortium and has received
funding from the Euratom research and training programme 2014-2018 under grant agreement No 633053.
The views and opinions expressed herein do not necessarily reflect those of the European Commission.
The authors gratefully acknowledge support from NAWI Graz
\red{and funding from the OeAD
under the grant agreement ``Wissenschaftlich-Technische
Zusammenarbeit mit der Ukraine'' No UA 06/2015}.
\end{acknowledgments}


%

\end{document}